\documentclass[10pt,twocolumn,twoside,journal]{IEEEtran}
\usepackage{subfigure}
\usepackage{setspace}
\usepackage{amsmath}
\usepackage{amssymb}
\usepackage{amsfonts}
\usepackage{amscd}
\usepackage{mathrsfs}
\usepackage[final]{graphicx}
\usepackage{graphicx}
\usepackage{psfrag}
\usepackage{url}
\usepackage{textcomp}
\usepackage{multirow}
\usepackage{cite}
\DeclareMathOperator*{\argmin}{argmin}

\newcommand{\Define}{\stackrel{\triangle}{=}}

\begin{document}

\markboth{Accepted in EURASIP JOURNAL ON ADVANCES IN SIGNAL PROCESSING:
Spl. Iss. on Multiuser MIMO Transmission with Limited Feedback...}{P. 
Ubaidulla and A. Chockalingam: \MakeLowercase{\textit{}} 
Robust THP Transceiver Designs for Multiuser MIMO Downlink
with Imperfect CSIT}

\title{Robust THP Transceiver Designs for Multiuser MIMO Downlink
with Imperfect CSIT}
\author{P. Ubaidulla and A. Chockalingam \IEEEmembership{Senior 
Member, IEEE}
\thanks{This work in part was presented in IEEE Wireless Communication 
and Networking Conference (WCNC'09), Budapest, Hungary, April 2009.
The authors are with the Department of Electrical Communication 
Engineering, Indian Institute of Science, Bangalore-560012, India.
E-mail: \{ubaidulla,achockal\}@ece.iisc.ernet.in}
}

\markboth{Accepted in EURASIP JOURNAL ON ADVANCES IN SIGNAL PROCESSING: 
Spl. Iss. on Multiuser MIMO Transmission with Limited Feedback...}
{P. Ubaidulla and A. Chockalingam: \MakeLowercase{\textit{}} 
Robust THP Transceiver Designs for Multiuser MIMO Downlink
with Imperfect CSIT}

\maketitle

\begin{abstract}
In this paper, we present robust joint non-linear transceiver designs
for multiuser multiple-input multiple-output (MIMO) downlink in the
presence of imperfections in the channel state information at the
transmitter (CSIT). The base station (BS) is equipped with multiple
transmit antennas, and each user terminal is equipped with {\em one or
more} receive antennas. The BS employs Tomlinson-Harashima precoding
(THP) for inter-user interference pre-cancellation at the transmitter.
We consider robust transceiver designs that jointly optimize the
transmit THP filters and receive filter for two models of CSIT errors.
The first model is a stochastic error (SE) model, where the CSIT error
is Gaussian-distributed. This model is applicable when the CSIT error is
dominated by channel estimation error. In this case, the proposed robust
transceiver design seeks to minimize a stochastic function of the sum
mean square error (SMSE) under a constraint on the total BS transmit
power. We propose an iterative algorithm to solve this problem. The
other model we consider is a norm-bounded error (NBE) model, where
the CSIT error can be specified by an uncertainty set. This model is
applicable when the CSIT error is dominated by quantization errors.
In this case, we consider a worst-case design. For this model, we
consider robust  $i)$ minimum SMSE, $ii)$ MSE-constrained, and
$iii)$ MSE-balancing transceiver designs. We propose iterative
algorithms to solve these problems, wherein each iteration involves
a pair of semi-definite programs (SDP). Further, we consider an
extension of the proposed algorithm to the case with per-antenna
power constraints. We evaluate the robustness of the proposed
algorithms to imperfections in CSIT through simulation, and show
that the proposed robust designs outperform non-robust designs as
well as robust linear transceiver designs reported in the recent
literature.
\end{abstract}


\section{Introduction}
\label{sec1}
Multiuser multiple-input multiple-output (MIMO) wireless communication
systems have attracted considerable interest due to  their potential to
offer the benefits of spatial diversity and increased capacity
\cite{tse06},\cite{bolc06}. Multiuser interference limits the performance
of such multiuser systems. To realize the potential of such systems in
practice, it is important to devise methods to reduce the multiuser
interference. Transmit-side processing at the base station (BS) in the
form of precoding has been studied widely as a means to reduce the multiuser
interference \cite{bolc06}. Several studies on linear precoding and
non-linear precoding (e.g., Tomlinson-Harashima precoder (THP)) have been
reported in the literature \cite{spenc04},\cite{kusum05}. Joint design of
both transmit precoder and receive filter can result in improved performance.
Transceiver designs that jointly optimize precoder/receive filters for
multiuser MIMO downlink with different performance criteria have been
widely reported in the literature \cite{doost05_2, bande06, zhang05,
shi07, mezgh06, mezgh06_2,shi08_c}. An important criterion that has been
frequently used in such designs is the sum mean square error (SMSE)
\cite{bande06, zhang05, shi07, mezgh06}. Iterative algorithms that
minimize SMSE with a constraint on total BS transmit power are reported
in \cite{bande06},\cite{zhang05}. These algorithms are not guaranteed to
converge to the global minimum. Minimum SMSE transceiver designs based on
uplink-downlink duality, which are guaranteed to converge to the global
minimum, have been proposed in \cite{shi07},\cite{mezgh06}. Non-linear
transceivers, though more complex, result in improved performance compared
to linear transceivers. Studies on non-linear THP transceiver design
have been reported in the literature. An iterative THP transceiver design
minimizing weighted SMSE has been reported in \cite{mezgh06_2}.
References \cite{shi07} and \cite{shi08_c}, which primarily consider
linear transceivers, present THP transceiver optimizations also as
extensions. In \cite{doost05}, a THP transceiver design minimizing
total BS transmit power under SINR constraints is reported.

All the studies on transceiver designs mentioned above assume the
availability of perfect channel state information at the transmitter
(CSIT). However, in practice, the CSIT is usually imperfect due to
different factors like estimation error, feedback delay, quantization,
etc. The performance of precoding schemes is sensitive to such inaccuracies
\cite{jinda05}. Hence, it is of interest to develop transceiver designs
that are robust to errors in CSIT. Linear and non-linear transceiver
designs that are robust to imperfect CSIT in multiuser multi-input
single-output (MISO) downlink, where each user is equipped with only a
single receive antenna, have been studied
\cite{hunge04, shen07_d, bigue04, payar07, vucic09_a, sheno_b}. Recently,
robust {\em linear} transceiver designs for multiuser MIMO downlink (i.e.,
each user is equipped with more than one receive antenna) based on the
minimization of the total BS transmit power under individual user MSE
constraints and MSE-balancing have been reported in \cite{vucic08}.
However, robust transceiver designs for
{\em non-linear} THP in multiuser {\em MIMO} with imperfect CSIT, to our
knowledge, have not been reported so far, and this forms the main focus
of this paper.

\begin{figure*}
\centering
\includegraphics[width=5.00in,height=2.25in]{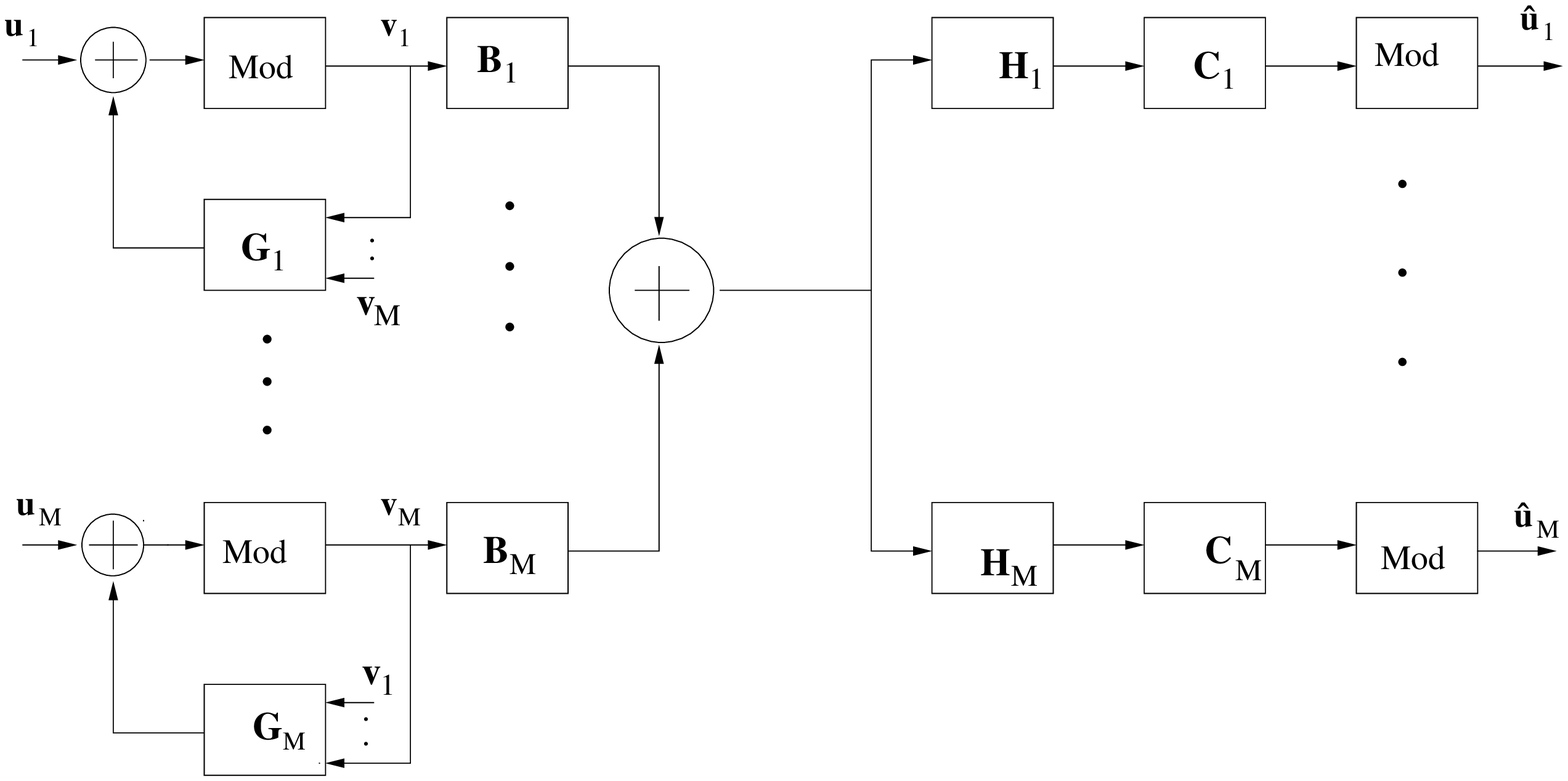}
\caption{
Multiuser MIMO downlink system model with Tomlinson-Harashima precoding.}
\label{fig1}
\end{figure*}

In this paper, we consider robust THP transceiver designs for multiuser
MIMO downlink in the presence of imperfect CSIT. We consider two widely
used models for the CSIT error \cite{models}, and propose robust THP
transceiver
designs suitable for these models. First, we consider a stochastic error
(SE) model for the CSIT error, which is applicable in TDD systems where
the error is mainly due to inaccurate channel estimation (in TDD,
the channel gains on uplink and downlink are highly correlated, and so
the estimated channel gains at the transmitter can be used for precoding
purposes). The error in this model is assumed to follow a Gaussian
distribution. In this case, we adopt a statistical approach, where
the robust transceiver design is based on minimizing the SMSE averaged
over the CSIT error. To solve this problem, we propose an iterative
algorithm, where each iteration involves solution of two sub-problems,
one of which can be solved analytically and the other is formulated as
a second order cone program (SOCP) that can be solved efficiently.
Next, we consider a norm-bounded error (NBE) model for the CSIT error,
where the error is specified in terms of uncertainty set of known size.
This model is suitable for FDD systems where the errors are mainly due
to quantization of the channel feedback information \cite{payar07}. In
this case, we adopt a min-max approach to the robust design, and propose
an iterative algorithm which involves the solution of semi-definite
programs (SDP). For the NBE model, we consider three design problems:
$i$) robust minimum SMSE transceiver design $ii$) robust MSE-constrained
transceiver design, and $iii$) robust MSE-balancing transceiver design.
We also consider the extension of the robust designs to incorporate
per-antenna power constraints. Simulation results show that the proposed
algorithms are robust to imperfections in CSIT, and they perform better
than non-robust designs as well as robust linear designs reported recently
in the literature.

The rest of the paper is organized as follows. The system model and
the CSIT error models are presented in Section \ref{sec2}. The
proposed robust THP transceiver design for SE model of CSIT error
is presented in Section \ref{sec3}. The proposed robust transceiver
designs for NBE model of CSIT error are presented in Section \ref{sec4}.
Simulation results and performance comparisons are presented in Section
\ref{sec5}. Conclusions are presented in Section \ref{sec6}.

\section{System Model}
\label{sec2}
We consider a multiuser MIMO downlink, where a BS communicates with
$M$ users on the downlink. The BS employs Tomlinson-Harashima precoding
for inter-user interference pre-cancellation (see the system model in
Fig. \ref{fig1}). The BS employs $N_t$
transmit antennas and the $k$th user is equipped with $N_{r_k}$ receive
antennas, $1 \leq k \leq M$. Let ${\bf u}_k$ denote\footnote{We use the
following notation: Vectors are denoted by boldface lowercase
letters, and matrices are denoted by boldface uppercase letters. $[.]^T$,
$[.]^H$, and $[.]^{\dagger}$, denote transpose, Hermitian, and
pseudo-inverse operations, respectively. $[{\bf A}]_{ij}$ denotes the
element on the $i$th row and $j$th column of the matrix ${\bf A}$.
vec$(.)$ operator stacks the columns of the input matrix into one
column-vector. $\|\cdot\|_F$ denotes the Frobenius norm,
and ${\mathbb E}\{\cdot\}$ denotes expectation operator.
${\bf A}\succeq{\bf B}$ implies ${\bf A}-{\bf B}$ is positive
semi-definite.} the
$L_k \times 1$ data symbol vector for the $k$th user, where $L_k$,
$k=1,2,\cdots,M$, is the number of data streams for the $k$th user.
Stacking the data vectors for all the users, we get the global data
vector ${\bf u}=[{\bf u}_1^T,\cdots,{\bf u}_{M}^T]^T.$ The output
of the $k$th user's modulo operator at the transmitter is denoted
by ${\bf v}_k$. Let ${\bf B}_k \in {\mathbb C}^{N_t \times {L_k}}$
represent the precoding matrix for the $k$th user. The global
precoding matrix ${\bf B}=[{\bf B}_1, {\bf B}_2,\cdots, {\bf B}_{M}]$.
The transmit vector is given by
\begin{eqnarray}
\label{e1}
{\bf x} & = & {\bf B}{\bf v},
\end{eqnarray}
where ${\bf v}=[{\bf v}_1^T,\cdots,{\bf v}_{M}^T]^T.$
The feedback filters are given by
\begin{eqnarray}
\hspace{-6mm}
{\bf G}_k & \hspace{-1mm} = & \hspace{-1mm} \left[{\underline{\bf G}}_{k,1} \cdots
{\underline{\bf G}}_{k,k-1}
\quad {\bf 0}_{L_k \times \sum_{j=k}^{M}L_j}\right],\,\,\, 1 \leq k \leq M,
\end{eqnarray}
where ${\underline{\bf G}}_{kj} \in {\mathbb C}^{L_{k} \times L_{j}}$,
perform the interference
pre-subtraction. We consider only inter-user interference pre-subtraction.
When THP is used, both the transmitter and the receivers employ the
modulo operator,
$\text{ Mod}(\cdot)$. For a complex number $x$, the modulo operator performs
the following operation
\begin{eqnarray}
\label{mdlo}
\hspace{-5mm}
\text{ Mod}(x) & = & x-a \left\lfloor \frac{\Re(x)}{a}+\frac{1}{2}\right\rfloor -{\textbf j}\, a \left\lfloor \frac{\Im(x)}{a}+\frac{1}{2}\right\rfloor,
\end{eqnarray}
where
${\textbf j}=\sqrt{-1}$, and $a$ depends on the constellation \cite{fisch_bk02}. For a vector argument
${\bf x}=[x_1\,\, x_2\,\, \cdots\, x_N]^T$,
\begin{eqnarray}
\hspace{-2mm}
\text{ Mod}({\bf x}) & = & \left[\text{Mod}(x_1)\,\, \text{Mod}(x_2) \cdots \text{ Mod}(x_N)\right]^T.
\end{eqnarray}
The vectors ${\bf u}_k$ and ${\bf v}_k$ are related as
\begin{eqnarray}
\label{vandu}
{\bf v}_k & = & \text{Mod}\Big({\bf u}_k-\sum_{j=1}^{k-1}{\underline{\bf G}}_{k,j}{\bf v}_j\Big).
\end{eqnarray}
The $k$th component of the transmit vector ${\bf x}$ is transmitted
from the $k$th transmit antenna. Let ${\bf H}_k$ denote the
$N_{r_k}\times N_t$ channel matrix of the $k$th user. The overall
channel matrix is given by
\begin{eqnarray}
{\bf H} & = & \left[{\bf H}_1^T \quad {\bf H}_2^T\cdots{\bf H}_{M}^T\right]^T.
\end{eqnarray}
The received signal
vectors are given by
\begin{eqnarray}
\label{e2}
{\bf y}_k&=&{\bf H}_k{\bf B}{\bf v}+{\bf n}_k, \,\,\,\,\,
1 \leq k \leq M.
\end{eqnarray}
The $k$th user estimates its data vector as
\begin{eqnarray}
\label{e3}
\hspace{-4mm}
{\widehat {\bf u}}_k&=&\left({\bf C}_k{\bf y}_k\right) \hspace{-3mm} \mod a \nonumber \\
&=& \left({\bf C}_k{\bf H}_k{\bf B}{\bf v}+{\bf C}_k{\bf n}_k\right) \hspace{-2mm} \mod a, \,\,\,\,\,\, 1\leq k \leq M,
\end{eqnarray}
where ${\bf C}_k$ is
the $L_k \times N_{r_k}$ dimensional receive filter of the $k$th user,
and ${\bf n}_k$ is the zero-mean noise vector with ${\mathbb E}\{{\bf
  n}_k{\bf n}_k^H\}=\sigma_n^2{\bf I}$.
Stacking the estimated vectors of all users, the global estimate vector
can be written as
\begin{eqnarray}
\label{e4}
{\widehat {\bf u}}&=&\left({\bf C}{\bf H}{\bf B}{\bf v}+{\bf C}{\bf n}\right) \hspace{-1mm} \mod a,
\end{eqnarray}
where ${\bf C}$ is a block diagonal matrix with ${\bf C}_k, \, 1\leq k\leq M$
on the diagonal, and ${\bf n}=[{\bf n}_1^T,\cdots,{\bf n}_{M}^T]^T$. The
global receive matrix ${\bf C}$ has block diagonal structure as the
receivers are non-cooperative. Neglecting the modulo loss, and assuming
${{\mathbb E}\{\bf v}_k{\bf v}_k^H\}={\bf I}$, we can write MSE between
the symbol vector ${\bf u}_k$ and the estimate ${\widehat{\bf u}}_k$ at
the $k$th user as \cite{mezgh06_2}
\begin{eqnarray}
\label{e5}
\epsilon_k & \hspace{-1mm} = & \hspace{-1mm}{\mathbb E}\{\|{\widehat{\bf u}}_k-{\bf u}_k\|^2\} \nonumber \\
&\hspace{-1mm} = &\hspace{-1mm}\mbox{tr}\Big[({\bf C}_k{\bf H}_k{\bf B}-\bar{{\bf G}}_k)({\bf C}_k{\bf H}_k {\bf B}-\bar{{\bf G}}_k)^H + \sigma_n^2{\bf C}_k{\bf C}_k^H\Big], \nonumber \\ 
& & \hspace{5cm} 1\leq k \leq M,
\end{eqnarray}
where
${\bar{\bf G}}_k = \left[{\underline{\bf G}}_{k,1} \cdots
{\underline{\bf G}}_{k,k-1} \quad {\bf I}_{L_k,L_k}\quad
{\bf 0}_{L_k \times \sum_{j=k+1}^{M}L_j}\right].$

\subsection{CSIT Error Models}
\label{sec2b}
We consider two models for the CSIT error. In both the models, the
true channel matrix of the $k$th user, ${\bf H}_k$, is represented as
\begin{eqnarray}
\label{e6aa}
{\bf H}_k&=&{\widehat{{\bf H}}}_k + {\bf E}_k,\quad 1 \leq k \leq M,
\end{eqnarray}
where ${\widehat{{\bf H}}}_k$ is the CSIT of the $k$th
user, and ${\bf E}_k$ is the CSIT error matrix.
The overall channel matrix can be written as
\begin{eqnarray}
\label{e6bb}
{\bf H}&=&{\widehat{\bf H}} + {\bf E},
\end{eqnarray}
where
${\widehat{\bf H}}=[{\widehat{\bf H}}_1^T \quad {\widehat{\bf H}}_2^T \cdots {\widehat{\bf H}}_{M}^T]^T$, and  ${\bf E}=[{\bf E}_1^T \quad {\bf E}_2^T \cdots{\bf E}_{M}^T]^T$.
In a stochastic error (SE) model, ${\bf E}_k$ is the channel estimation
error matrix. The error matrix ${\bf E}_k$ is assumed to be Gaussian
distributed with zero mean and
${\mathbb E}\{{\bf E}_k{\bf E}_k^{H}\}=\sigma^{2}_{E}{\bf I}_{N_{r_k}N_{r_k}}$.
This statistical model is suitable for systems with uplink-downlink
reciprocity. We use this model in Sec. \ref{sec3}.
An alternate error model is a norm-bounded error (NBE) model, where
\begin{eqnarray}
\label{errnorm}
\|{\bf E}_k\|_F & \leq & \delta_k, \quad 1 \leq k \leq M,
\end{eqnarray}
or, equivalently, the true channel ${\bf H}_k$ belongs to the
uncertainty set ${\cal R}_k$ given by
\begin{eqnarray}
\label{uset}
\hspace{-7mm}
{\cal R}_k &\hspace{-1.5mm} = & \hspace{-1.5mm} \{{\boldsymbol \zeta}\big |
{\boldsymbol \zeta}={\widehat{\bf H}}_k+{\bf E}_k,
\|{\bf E}_k\|_F \leq \delta_k\}, \quad 1 \leq k \leq M,
\end{eqnarray}
where $\delta_k$ is the CSIT {\em uncertainty size}. This model is
suitable for systems where quantization of CSIT is involved
\cite{payar07}. We use this model in Sec. \ref{sec4}.

\section{Robust Transceiver Design with Stochastic CSIT Error}
\label{sec3}
In this section, we propose a transceiver design that minimizes SMSE
under a constraint on total BS transmit power and is robust in the
presence of CSIT error, which is assumed to follow the SE model.
This involves the joint design of the precoder ${\bf B}$, feedback
filter ${\bf G}$, and receive filter ${\bf C}$. When ${\bf E}$, the
CSIT error matrix, is a random matrix, the SMSE is a random variable.
In such cases, where the objective function to be minimized is a
random variable, we can consider the minimization of the expectation
of the objective function. In the present problem, we adopt this
approach. Further, the computation of the expectation of SMSE with
respect to ${\bf E}$ is simplified as ${\bf E}$ follows Gaussian
distribution. Following this approach, the robust transceiver design
problem can be written as
\begin{eqnarray}
\label{e9}
\min_{{\bf B},{\bf C},{\bf G}} & & {\mathbb E}_{\bf E}\{\texttt{smse}\} \\
\mbox{subject to} & & \text{Tr}\left({\bf B}{\bf B}^H\right) \leq P_{max}, \nonumber
\end{eqnarray}
where $P_{max}$ is the limit on the total BS transmit power, and
minimization over ${\bf B},{\bf C},{\bf G}$ implies minimization over
${\bf B}_i,{\bf C}_i,{\bf G}_i,\,\,1 \leq i \leq M$.
Incorporating the imperfect CSIT,
${\bf H}={\widehat {\bf H}} +{\bf E}$, in (\ref{e5}), the
SMSE can be written as
\begin{eqnarray}
\label{e10}
\hspace{-2mm}
\texttt{smse}&=&
{\mathbb E}\{\|{\widehat {\bf u}}-{\bf u}\|^2\} \nonumber\\
&\hspace{-23mm}=& \hspace{-13mm}\sum_{k=1}^M\mbox{tr}\Big[({\bf C}_k({\widehat{\bf H}}_k+{\bf E}_k)
{\bf B}-\bar{{\bf G}}_k)({\bf C}_k({\widehat{\bf H}_k} + {\bf E}_k)
{\bf B}-\bar{{\bf G}}_k)^H \nonumber \\
& & +\, \sigma_n^2{\bf C}_k{\bf C}_k^H\Big].
\end{eqnarray}
Averaging the $\texttt{smse}$ over ${\bf E}$, we write the new
objective function as
\begin{eqnarray}
\label{e11}
\mu& \Define &\mathbb{E}_{\bf E}\{\texttt{smse}\} \nonumber\\
&=&\sum_{k=1}^M\mbox{tr}\Big[({\bf C}_k{\widehat{\bf H}}_k{\bf B}
 - \bar{{\bf G}}_k)({\bf C}_k{\widehat{\bf H}_k}{\bf B}
- \bar{{\bf G}}_k)^H \nonumber \\
& & + \, \big(\sigma_{E}^2\mbox{tr}({\bf B}{\bf B}^H)
+\sigma_n^2\big) {\bf C}_k{\bf C}_k^H\Big].
\end{eqnarray}
Using the objective function $\mu$, the robust transceiver design
problem can be written as
\begin{eqnarray}
\label{e11a}
\min_{{\bf B},{\bf C},{\bf G}} & &\mu \\
\mbox{subject to} & & \|{\bf B}\|_F^2 \leq P_{max}. \nonumber
\end{eqnarray}
From \eqref{e11}, we observe that $\mu$ is not
jointly convex in ${\bf B}$, ${\bf G}$, and ${\bf C}$.
However, it is convex in ${\bf B}$ and
${\bf G}$ for a fixed value of ${\bf C}$, and vice versa. So, we propose
an iterative algorithm in order to solve the problem in \eqref{e11a},
where each iteration involves the solution of a sub-problem which either
has an analytic solution or can be formulated as a convex optimization
program.

\subsection{Robust Design of ${\bf G}$ and ${\bf C}$ Filters}
\label{sec31}
Here, we consider the design of robust feedback and receive filters,
${\bf G}$ and ${\bf C}$, that minimizes the \texttt{smse} averaged
over ${\bf E}$. For a given ${\bf B}$ and ${\bf C}_k$, as we can see
from (\ref{e11}), the optimum feedback filter
${\underline{\bf G}}_{k,j},\,\,1 \leq k \leq M, \,\, j < k,$ is given by
\begin{eqnarray}
\label{optf}
{\underline{\bf G}}_{k,j} & = & {\bf C}_k{\widehat{\bf H}}_k{\bf B}_j.
\end{eqnarray}
Substituting the optimal ${\underline{\bf G}}_{k,j}$ given above in
\eqref{e11}, the objective function can be written as
\begin{eqnarray}
\label{eqmu1}
\mu&\hspace{-0mm}=&\hspace{-0mm}\sum_{k=1}^M\mbox{tr}\Big[({\bf C}_k{\widehat{\bf H}}_k{\bf B}_k - {\bf I})({\bf C}_k{\widehat{\bf H}_k}{\bf B}_k - {\bf I})^H \nonumber \\
&\hspace{-0mm} & \hspace{-0mm} +\sum_{j=k+1}^M ({\bf C}_k{\widehat{\bf H}}_k{\bf B}_j) ({\bf C}_k{\widehat{\bf H}}_k{\bf B}_j)^H
\nonumber \\ 
&& + \, \big(\sigma_{E}^2\mbox{tr}({\bf B}{\bf B}^H)+\sigma_n^2\big) {\bf C}_k{\bf C}_k^H\Big].
\end{eqnarray}
In order to compute the optimum receive filter, we differentiate
(\ref{eqmu1}) with respect to ${\bf C}_k,\,\, 1 \leq k \leq M,$ and
set the result to zero. We get
\begin{eqnarray}
\label{e15}
\hspace{-5mm}
{\bf B}_k^H{\widehat{\bf H}}_k^H&=&{\bf C}_k\bigg({\widehat{\bf H}}_k\Big(\sum_{j=k+1}^{M}{\bf B}_j{\bf B}_j^H\Big){\widehat{\bf H}}_k^H \nonumber \\
& & + \left(\sigma_n^2+\sigma_{ E}^2\|{\bf B}\|_F^2\right){\bf I}\bigg),\quad 1 \leq k \leq M.
\end{eqnarray}
From the above equation, we get
\begin{eqnarray}
\label{e16}
{\bf C}_k&=&{\bf B}_k^H{\bf  H}_k^H \bigg({\widehat{\bf H}}_k\Big(\sum_{j=k+1}^{M}{\bf B}_j{\bf B}_j^H\Big){\widehat{\bf H}}_k^H \nonumber \\
& & + \,\big(\sigma_n^2+\sigma_{ E}^2\|{\bf B}\|_F^2\big){\bf I}\bigg)^{-1}, \quad 1 \leq k \leq M.
\end{eqnarray}
We observe that the expression for the robust receive filter in \eqref{e16}
is similar to the standard MMSE receive filter, but with an additional
factor that account for the CSIT error. In case of perfect CSIT,
$\sigma_{E}=0$ and the expression in \eqref{e16} reduces to the MMSE
receive filters in \cite{mezgh06_2}, \cite{doost05}.

\vspace{-3mm}
\subsection{Robust Design of ${\bf B}$ Filter}
\label{sec32}
Having designed the feedback and receive filter matrices, ${\bf G}$
and ${\bf C}$, for a given precoder matrix ${\bf B}$, we now present the
design of the robust precoder matrix for given feedback and receive filter
matrices. Towards this end, we express the robust transceiver design
problem in \eqref{e11a} as
\begin{eqnarray}
\label{vecpr}
\hspace{-6mm}
\min_{{\bf b},{\bf c},{\bf g}} & & \hspace{-3mm} \sum_{k=1}^M\|{\bf D}_k{\widehat{\bf
    h}}_k-{\bar{\bf g}}_k\|^2+(\sigma_E^2\|{\bf b}\|^2
+\sigma_n^2)\|{\bf c}_k\|^2  \\
\hspace{-6mm}
\text{subject to} & &\hspace{-3mm}\|{\bf b}\|^2 \leq P_{max}, \nonumber
\end{eqnarray}
where
${\bf D}_k=({\bf B}^T\otimes{\bf C}_k)$,
${\widehat{\bf h}}_k=\mbox{vec}({\widehat {\bf H}}_k)$,
${\bf b}=\mbox{vec}({\bf B})$,
${\bf c}_k=\mbox{vec}({\bf C}_k)$,
${\bar {\bf g}_k}=\mbox{vec}({\bar{\bf G}}_k)$, and
${\bf h}_k=\mbox{vec} ({\bf H}_k)$.
Minimization over ${\bf b},{\bf c},{\bf g}$ denotes minimization over
${\bf b}_i,{\bf c}_i,{\bar{\bf g}}_i,\,\,\, 1 \leq i \leq M$. For given
${\bf C}$ and ${\bf G}$, the problem given above is a convex
optimization problem. The robust precoder design problem, given
${\bf C}$ and ${\bf G}$, can be written as
\vspace{-4mm}
\begin{eqnarray}
\label{vpr}
\min_{\bf b} & &\hspace{-3mm}\sum_{k=1}^M\|{\bf D}_k{\widehat{\bf h}}_k-{\bar{\bf g}}_k\|^2+\sigma_E^2\|{\bf b}\|^2\|{\bf c}_k\|^2 \nonumber \\
& & +\, \sigma_n^2\|{\bf c}_k\|^2 \\
\text{subject to} & &\|{\bf b}\|^2 \leq P_{max}. \nonumber
\end{eqnarray}
As the last term in \eqref{vpr} does not affect the optimum value of
${\bf b}$, we drop this term. Dropping this term and introducing the
dummy variables $t_k,r_k,\,\,1 \leq k \leq M$, the problem in (\ref{vpr})
can be formulated as the following convex optimization problem:
\begin{eqnarray}
\label{e12}
\min_{{\bf b},\{t_i\}_1^M,\{r_i\}_1^M} & &\sum_{k=1}^M
t_k+\sigma_E\|{\bf c}_k\|^2r_k \\
\mbox{subject to} & & \|{\bf D}_k{\widehat {\bf h}}_k-
{\bar{\bf g}}_k\|^2\leq t_k,\nonumber\\
&&\|{\bf b}\|^2   \leq r_k, \nonumber\\
& & r_k \leq P_{max}, \,\,1 \leq k \leq M. \nonumber
\end{eqnarray}
The constraints in the above optimization problem are rotated second
order cone constraints \cite{boyd04}. Convex optimization problems
like that in \eqref{e12} can be efficiently solved using
interior-point methods \cite{sturm99,boyd04}.

\subsection{Iterative Algorithm to Solve (\ref{e9})}
Here, we present the proposed iterative algorithm for the minimization
of the SMSE averaged over ${\bf E}$ under total BS transmit power
constraint. In each iteration, the computations presented in subsections
\ref{sec31} and \ref{sec32} are performed. In the $(n+1)$th iteration,
the value of ${\bf B}$, denoted by ${\bf B}^{n+1}$, is the solution to
the following problem
\begin{eqnarray}
\label{e20}
{\bf B}^{n+1}&=&\argmin_{{\bf B}:\text{Tr}({\bf B}{\bf B}^H) \leq P_{max}} \mu({\bf
 B},{\bf C}^n,{\bf G}^n),
\end{eqnarray}
which is solved in the previous subsection.
Having computed ${\mathbf B}^{n+1}$, ${\mathbf C}^{n+1}$ is the
solution to the following problem:
\begin{eqnarray}
\label{e21}
{\bf C}^{n+1}&=&\argmin_{{\bf C}} \mu({\bf  B}^{n+1},{\bf C},{\bf G}^n),
\end{eqnarray}
and its solution is given in (\ref{e16}). Having computed ${\mathbf B}^{n+1}$
and ${\mathbf C}^{n+1}$,  ${\bf G}^{n+1}$ is the solution to the following
problem:
\begin{eqnarray}
\label{e21a}
{\bf G}^{n+1}&=&\argmin_{{\bf G}} \mu({\bf  B}^{n+1},{\bf
  C}^{n+1},{\bf G}),
\end{eqnarray}
and its solution is given in (\ref{optf}).
As the objective function in (\ref{e11}) is monotonically decreasing
after each iteration and is lower bounded, convergence is guaranteed.
The iteration is terminated when the norm of the difference in the
results of consecutive iterations are below a threshold or when the
maximum number of iterations is reached. We note that the proposed
algorithm is not guaranteed to converge to the global minimum.

\section{Robust Transceiver Designs with Norm-Bounded CSIT Error}
\label{sec4}
When the receivers quantize the channel estimate and send the CSI
to the transmitter through a low-rate feedback channel, we can model
the error in CSI at the transmitter by the NBE model \cite{payar07}.
In such cases, it is appropriate to consider the min-max  design, where
the worst-case value of the objective function is minimized. In this
section, we address robust transceiver designs in the presence of a
norm-bounded CSIT error. Specifically, we consider $i$) a robust SMSE
transceiver design, $ii$) a robust MSE-constrained transceiver design,
and $iii)$ a robust MSE-balancing (min-max fairness) design.

\subsection{Robust SMSE Transceiver Design}
\label{sec41}
Here, we consider a min-max design, wherein the design seeks to minimize
the worst case SMSE under a total BS transmit power constraint. This
problem can be written as
\begin{eqnarray}
\label{enbe}
\min_{{\bf B},{\bf C},{\bf G}}  & & \max_{{\bf E}_k:\|{\bf E}_k\| \leq \delta_k, \forall k}\texttt{smse}({{\bf B},{\bf C},{\bf G},{\bf E}}) \\
\mbox{subject to}  &&\mbox{tr}({\bf B}{\bf B}^H) \leq P_{max}. \nonumber
\end{eqnarray}
The above problem deals with the case where the true channel, unknown to
the transmitter, may lie anywhere in the uncertainty region. In order to
ensure, a priori, that MSE constraints are met for the actual channel,
the precoder should be so designed that the constraints are met for all
members of the uncertainty set. This, in effect, is a semi-infinite
optimization problem \cite{benta07}, which in general is intractable.
We show, in the following, that an appropriate transformation makes
the problem in (\ref{enbe}) tractable. We note that the problem in
(\ref{enbe}) can be written as
\begin{eqnarray}
\label{enbe1}
\hspace{-6mm}
\min_{{\bf b},{\bf c},{\bf g},t}  && \sum_{k=1}^M t_k \\
\hspace{-6mm}
\mbox{subject to}&&\|{\bf D}_k({\widehat{\bf h}}_k+{\bf e}_k)-{\bar{\bf g}}_k\|^2+\sigma_n^2\|{\bf c}_k\|^2 \leq t_k, \\ \nonumber 
& & \hspace{2cm} \forall\,\,\|{\bf e}_k\| \leq \delta_k, \,1 \leq k \leq M, \nonumber \\
&&\|{\bf b}\|^2 \leq P_{max}, \nonumber
\end{eqnarray}
where ${\bf e}_k=\text{vec}({\bf E}_k)$. The first constraint in
(\ref{enbe1}) is convex in ${\bf B}$ and ${\bar{\bf G}}_k$ for a fixed value
of ${\bf C}_k$ and vice versa, but not jointly convex in ${\bf B}$,
${\bar{\bf G}}_k$ and ${\bf C}_k$. Hence, to design the transceiver, we
propose an iterative algorithm, wherein the optimization is performed
alternately over $\{{\bf B},{\bf G}\}$ and $\{{\bf C}\}$.

\subsubsection{Robust Design of ${\bf B}$ and ${\bf G}$ Filters}
For the design of the precoder matrix ${\bf B}$ and the feedback filter
${\bf G}$ for a fixed value of ${\bf C}$, the second term in the left
hand side of the first constraint in (\ref{enbe1}) is not relevant, and
hence we drop this term. Invoking the Schur Complement Lemma \cite{horn85},
and dropping the second term, we can write the constraint in (\ref{enbe1})
as the following linear matrix inequality (LMI):
\begin{eqnarray}
\label{sdp1}
\hspace{-7.5mm}
\left[\begin{array}{cc}
t_k & \left[{\bf D}_k({\widehat{\bf h}}_k+{\bf e}_k)-{\bar{\bf g}}_k\right]^H \\
\left[{\bf D}_k({\widehat{\bf h}}_k+{\bf e}_k)-{\bar{\bf g}}_k\right] & {\bf I}
\end{array}\right] & \hspace{-3.5mm} \succeq & \hspace{-3mm} {\bf 0}.
\end{eqnarray}
Hence, the robust precoder and feedback filter design problem, for a
given ${\bf C}$, can be written as
\begin{eqnarray}
\label{enbe1a}
\min_{{\bf B},{\bf G},t}  && \sum_{k=1}^M t_k\\
\mbox{subject to}&&  \left[\begin{array}{cc}
 t_k & \left[{\bf D}_k{{\bf h}}_k-{\bar{\bf g}}_k\right]^H \nonumber\\
\left[{\bf D}_k{{\bf h}}_k-{\bar{\bf g}}_k\right] & {\bf I}
 \end{array}\right] \, \succeq \, {\bf 0},  \nonumber\\
&& \forall \, \|{\bf e}_k\| \leq \delta_k,\quad 1\leq k \leq M, \nonumber\\
&&\|{\bf b}\| \leq \sqrt{P_{max}}, \nonumber
\end{eqnarray}
where ${\bf h}_k={\widehat{\bf h}}_k+{\bf e}_k$. From (\ref{sdp1}), the
first constraint in (\ref{enbe1a}) can be written as
\begin{eqnarray}
\label{nform}
{\bf A} & \succeq & {\bf P}^H{\bf X}{\bf Q}+{\bf Q}^H{\bf X}^H{\bf P},
\end{eqnarray}
where
\begin{eqnarray}
{\bf A} & = & \left[\begin{array}{cc}
 t_k & \left[{\bf D}_k{\widehat{\bf h}}_k-{\bar{\bf g}}_k\right]^H \\
\left[{\bf D}_k{\widehat{\bf h}}_k-{\bar{\bf g}}_k\right] & {\bf I}
 \end{array}\right],
\end{eqnarray}
${\bf P}=\left[{\bf 0} \quad {\bf D}_k^H\right]$, ${\bf X}={\bf e}_k$, and ${\bf Q}=-\left[1
  \quad {\bf 0}\right].$
Having reformulated the constraint as in (\ref{nform}), we can
invoke the following Lemma \cite{eldar05} to solve the problem in
(\ref{enbe1a}).\\
\hspace{2mm}{\em Lemma 1:} Given matrices ${\bf P}$, ${\bf Q}$, ${\bf
  A}$ with ${\bf A}={\bf A}^H$,
\begin{eqnarray}
\label{eld0}
{\bf A} & \succeq & {\bf P}^H{\bf X}{\bf Q}+{\bf Q}^H{\bf X}^H{\bf P}, \quad \forall \, {\bf X}:\|{\bf X}\|\leq \rho,
\end{eqnarray}
if and only if $\exists \, \lambda \geq 0$ such that
\begin{eqnarray}
\label{eld}
\left[\begin{array}{cc}
{\bf A}-\lambda{\bf Q}^H{\bf Q} & -\rho{\bf P}^H \\
-\rho{\bf P} & \lambda{\bf I}
 \end{array}\right] & \succeq & {\bf 0}\,.
\end{eqnarray}
\hfill$\square$

Applying Lemma 1, we can formulate the robust precoder design problem
as the following convex optimization problem:
\begin{eqnarray}
\label{sd11}
\min_{{\bf B},{\bf G},t,\beta}  && \sum_{k=1}^M t_k\\
\mbox{subject to}&&  {\bf M}_k\succeq{\bf 0},\,\,\beta_k \geq 0,  \,\,\forall k,\nonumber\\
&&\|{\bf b}\| \leq \sqrt{P_{max}}, \nonumber
\end{eqnarray}
where
\begin{eqnarray}
\label{mk}
\hspace{-7mm}
{\bf M}_k & \hspace{-1.5mm} = & \hspace{-1.5mm} \left[\begin{array}{ccc}
t_k-\beta_k & ({\bf D}_k{\widehat {\bf h}}_k-{\bar{\bf g}}_k)^H & {\bf
  0}\\
 ({\bf D}_k{\widehat {\bf h}}_k-{\bar{\bf g}}_k)  & {\bf I}  & -\delta_k{\bf
  D}_k\\
{\bf 0} & -\delta_k{\bf D}_k^H & \beta_k{\bf I}\end{array}\right]\hspace{-1mm}.
\end{eqnarray}

\subsubsection{Robust Design of Filter Matrix ${\bf C}$}
In the previous subsection, we considered the design of the ${\bf B}$
and ${\bf G}$ matrices for a fixed ${\bf C}$. Here, we consider the
robust design ${\bf C}$ for given ${\bf B}$ and ${\bf G}$. This design
problem can be written as
\begin{eqnarray}
\label{enber}
\min_{{\bf C},t}  && \sum_{k=1}^M t_k\\
\mbox{subject to}&&\|{\bf D}_k({\widehat{\bf h}}_k+{\bf e}_k)-{\bar{\bf
  g}}_k\|^2+\sigma_n^2\|{\bf c}_k\|^2 \leq
t_k,\nonumber\\
&& \forall\, \|{\bf E}_k\| \leq \delta_k,\quad 1 \leq k \leq M. \nonumber
\end{eqnarray}
Applying the Schur Complement Lemma, we can represent the first
constraint in (\ref{enber}) as
\begin{eqnarray}
\label{sdp2}
\hspace{-7mm}
\begin{bmatrix}
t_k & \begin{bmatrix}
{\bf D}_k({\widehat{\bf h}}_k+{\bf e}_k)-{\bar{\bf
     g}}_k\\
 \sigma_n{\bf c}_k\end{bmatrix}^H \\
\begin{bmatrix}
{\bf D}_k({\widehat{\bf h}}_k+{\bf e}_k)-{\bar{\bf g}}_k\\
\sigma_n{\bf c}_k\end{bmatrix} & {\bf I}
\end{bmatrix} & \hspace{-2.5mm} \succeq & \hspace{-2mm} {\bf 0}.
\end{eqnarray}
The second inequality in the above problem, like in the precoder
design problem, represents an infinite number of constraints. To make
the problem in \eqref{enber} tractable, we again invoke Lemma 1.
Following the same procedure as in the precoder design, starting with
(\ref{sdp2}), we can reformulate the robust receive filter design as the
following convex optimization problem:
\begin{eqnarray}
\label{sd12}
\min_{{\bf C},t,\lambda}  && \sum_{k=1}^M t_k\\
\mbox{subject to}&&  {\bf N}_k\succeq{\bf 0}, \,\,\,\forall k,\nonumber
\end{eqnarray}
where
\begin{eqnarray}
\label{mka}
\hspace{-6mm}
{\bf N}_k & \hspace{-1.5mm} = & \hspace{-1.5mm} \left[\begin{array}{ccc}
t_k-\lambda_k & \begin{bmatrix}
({\bf D}_k{\widehat {\bf h}}_k-{\bar{\bf g}}_k\\
  \sigma_n{\bf c}_k\end{bmatrix}^H & {\bf  0}\\
 \begin{bmatrix}{\bf D}_k{\widehat {\bf h}}_k-{\bar{\bf g}}_k \\
 \sigma_n{\bf c}_k\end{bmatrix}  &
        {\bf I}
 & -\delta_k{\bf \Gamma}_k\\
{\bf 0} & -\delta_k{\bf \Gamma}_k^H & \lambda_k{\bf I}\end{array}\hspace{-2mm}\right],
\end{eqnarray}
where ${\bf \Gamma}_k=\begin{bmatrix} {\bf D}_k\\ {\bf 0} \end{bmatrix}$.

\vspace{4mm}
\subsubsection{Iterative Algorithm to Solve (\ref{enbe})}
In the previous subsections, we described the design of ${\bf B}$
and ${\bf G}$ for a given ${\bf C}$, and vice versa. Here, we present
the proposed iterative algorithm for the minimization of the SMSE
under a constraint on the total BS transmit power, when the CSIT error
follows NBE model. The algorithm alternates between the optimizations
of the precoder/feedback filter and receive filter described
in the previous subsections. At the $(n+1)$th
iteration, the value of ${\bf B}$, denoted by ${\bf B}^{n+1}$, is the
solution to problem (\ref{sd11}), and hence satisfies the BS transmit
power constraint. Having computed ${\mathbf B}^{n+1}$,
${\mathbf C}^{n+1}$ is the solution to the problem in (\ref{sd12}). So
$J({\bf B}^{n+1},{\bf C}^{n+1})\leq J({\bf B}^{n+1},{\bf C}^{n})\leq J({\bf B}^{n},{\bf C}^{n})$,
where
\begin{eqnarray}
\label{defj}
J({\bf B},{\bf C}) & = & \max_{{\|{\bf E}_k\|<\delta_k, \forall k}} \{\texttt{smse}({\bf B},{\bf C},{\bf G},{\bf E})\}.
\end{eqnarray}
The monotonically decreasing nature of $J({\bf B}^n,{\bf C}^n)$,
together with the fact that $J({\bf B}^n,{\bf C}^n)$ is lower-bounded,
implies that the proposed algorithm converges to a limit as
$\mathop{n \to \infty}$. The iteration is terminated when the norm
of the difference in the results of consecutive iterations are below a
threshold or when the maximum number of iterations is reached.
This algorithm is not guaranteed to converge to the global minimum.

\vspace{2mm}
\subsubsection{Transceiver Design with Per-Antenna Power Constraints}
As each antenna at the BS usually has its own amplifier, it is
important to consider transceiver design with constraints on power
transmitted from each antenna. A precoder design for multiuser MISO
downlink with per-antenna power constraint with perfect CSIT was
considered in \cite{yu07}. Here, we incorporate per-antenna power
constraint in the proposed robust transceiver design. For this, only
the precoder matrix design (\ref{sd11}) has to be modified by including
the constraints on power transmitted from each antenna as given below:
\begin{eqnarray}
\label{sd11aa}
\min_{{\bf b}}  && \sum_{k=1}^M t_k\\
\mbox{subject to}&&  {\bf M}_k\succeq{\bf 0} \,\,\forall k,\nonumber\\
&& \|{\boldsymbol{\phi}}_k{\bf B}\|^2 \leq P_k, \quad 1 \leq k \leq M,\nonumber
\end{eqnarray}
where
${\boldsymbol{\bf \phi}}_k=[{\bf 0}_{1 \times k-1} \quad 1 \quad {\bf 0}_{1 \times N_t-k}]$.
The receive filter can be computed using (\ref{sd12}).

\subsection{Robust MSE-Constrained Transceiver Design}
\label{sec42}
Transceiver designs that satisfy QoS constraints are of interest. Such
designs in the context of multiuser MISO downlink with perfect CSI have been
reported in the literature \cite{wiese06},\cite{ schub04},\cite{schub05a}.
Robust linear precoder designs for MISO downlink with SINR constraints are
described in \cite{sheno07}. Here, we address the problem of robust THP
transceiver design for multiuser MIMO with MSE constraints in the presence
of CSI imperfections. THP designs are of interest because of their better
performance compared to the linear designs.

When the CSIT is perfect, the transceiver design under MSE constraints
can be written as
\begin{eqnarray}
\label{qos1}
\min_{{\bf B},{\bf G},{\bf C}} &&  \text{tr}({\bf B}{\bf B}^H) \\
\text{subject to} && \epsilon_k \leq \eta_k, \, 1 \leq k \leq M, \nonumber
\end{eqnarray}
where $\eta_k$ is the maximum allowed MSE at $k$th user
terminal. This problem can be written as the following
optimization problem:
\begin{eqnarray}
\label{mbaln1}
\min_{{\bf B},{\bf G},{\bf C},r} && r\\
\mbox{subject to}&&\|{\bf D}_k{\bf h}_k-{\bar{\bf
  g}}_k\|^2+\sigma_n^2\|{\bf c}_k\|^2 \leq
\eta_k, \,\,1 \leq k \leq M,\nonumber\\
&&\|{\bf b}\|^2 \leq r, \nonumber
\end{eqnarray}
where $r$ is a slack variable.
With the NBE model of imperfect CSI, the robust transceiver design with MSE
constraints can be written as
{\begin{eqnarray}
\label{mbaln2}
\min_{{{\bf b}},{{\bf g}},{{\bf c}},r} && r\\
\mbox{subject to}&&\|{{\bf D}}_k{{\bf h}}_k-
{{\bar{\bf g}}}_k\|^2+\sigma_n^2\|{{\bf c}}_k\|^2 \leq \eta_k, \nonumber \\ 
& & \hspace{2cm} \forall \, {{\bf h}}_k \in {\cal R}_k,\,1 \leq k \leq M, \nonumber\\
&&\|{{\bf b}}\|^2 \leq r. \nonumber
\end{eqnarray}
In the above problem, the true channel, unknown to the transmitter, may
lie anywhere in the uncertainty region.
The transceiver should be so designed that the constraints are met for
all members of the uncertainty set, ${\cal R}_k$. This again, in the
present form, is a semi-infinite optimization problem. In the following,
we present a transformation that makes the problem in (\ref{mbaln2})
tractable.

The optimization problem in \eqref{mbaln2} is not jointly convex in
${\bf b}$, ${\bf g}$, and ${\bf c}$. But, for fixed ${\bf c}$, it is
convex in ${\bf b}$ and ${\bf g}$, and vice versa. So, in order to solve
this problem, we propose an alternating optimization
algorithm\footnote{For the case of single antenna users (i.e., MISO),
a solution based on non-alternating approach is presented in
\cite{sheno_b}.}, wherein
each iteration solves two sub-problems. The first sub-problem given below,
involves the optimization over $\{{\bf b},{\bf g}\}$ for fixed ${\bf c}$.
\begin{eqnarray}
\label{s1def}
\min_{{{\bf b}},{{\bf g}},r} && r \\
\mbox{subject to}&&\|{{\bf D}}_k{{\bf h}}_k -
{{\bar{\bf g}}}_k\|^2+\sigma_n^2\|{{\bf c}}_k\|^2 \leq \eta_k, \nonumber \\ 
& & \hspace{2cm} \forall \, {{\bf h}}_k \in {\cal R}_k,\,1 \leq k \leq M, \nonumber\\
&&\|{{\bf b}}\|^2 \leq r. \nonumber
\end{eqnarray}
The second sub-problem involves optimization over $\{{\bf c}\}$ for
fixed $\{{\bf b},{\bf g}\}$, as given below
\begin{eqnarray}
\label{s2def}
\min_{{{\bf c}}, {s_1, \cdots, s_M}}&& s_k \\
\text{subject to} &&\|{{\bf D}}_k{{\bf h}}_k - {{\bar{\bf g}}}_k\|^2+\sigma_n^2\|{{\bf c}}_k\|^2 \leq s_k, \nonumber \\ 
& & \hspace{2cm} \forall \, {{\bf h}}_k \in {\cal R}_k, \,\,\, 1 \leq k \leq M, \nonumber
\end{eqnarray}
where $s_1,\cdots,s_M$ are slack variables.
The first sub-problem  can be expressed as a semi-definite program (SDP),
which is a convex optimization problem that can be solved efficiently
\cite{boyd04}. Towards this end, we reformulate the problem in \eqref{s1def}
as the following SDP:

\vspace{-2mm}
{\footnotesize
\begin{eqnarray}
\label{conref1}
\min_{{{\bf b}},{{\bf g}},r}  && \hspace{-5mm} r\\
\text{subject to} && \hspace{-5mm} \left[\begin{array}{cc}
 \eta_k & \begin{bmatrix}
{{\bf D}}_k\big({\widehat{\bf h}}_k+{\bf e}_k\big)-{{\bar{\bf
      g}}}_k\nonumber\\
\sigma_n{ {\bf c}}_k
\end{bmatrix}^H \nonumber\\
\begin{bmatrix}
{{\bf D}}_k\big({\widehat{\bf h}}_k+{\bf e}_k\big)-{{\bar{\bf
      g}}}_k\nonumber\\
\sigma_n{ {\bf c}}_k
\end{bmatrix} & {\bf I}
 \end{array}\right] \, \succeq \, {\bf 0},  \nonumber\\
&&\hspace{3.5cm} \forall \, \|{{\bf e}}_k\| \leq \delta_k,\quad 1 \leq k \leq M, \nonumber\\
&&\|{{\bf b}}\|<r\nonumber,
\end{eqnarray}
}

\vspace{-4mm}
\hspace{-4mm}
where $r$ is a slack variable. In the reformulation given above, we
have transformed the first constraint in \eqref{s1def} into an LMI
using the Schur Complement Lemma \cite{horn85}.
We can show that the LMI in \eqref{conref1} is equivalent to
\begin{eqnarray}
\label{nform22}
{\bf A} & \succeq & {\bf P}^H{\bf X}{\bf Q}+{\bf Q}^H{\bf X}^H{\bf P},
\end{eqnarray}
where
\begin{eqnarray}
{\bf A} & = & \left[\begin{array}{cc}
 \eta_k & \begin{bmatrix}
{{\bf D}}_k{\widehat{\bf h}}_k-{{\bar{\bf
      g}}}_k\nonumber\\
\sigma_n{ {\bf c}}_k
\end{bmatrix}^H \\
\begin{bmatrix}
{{\bf D}}_k{\widehat{\bf h}}_k-{{\bar{\bf
      g}}}_k\nonumber\\
\sigma_n{ {\bf c}}_k
\end{bmatrix} & {\bf I}
 \end{array}\right],
\end{eqnarray}
${\bf P}=\left[{\bf 0} \quad {\bf {\Gamma}}_k^H\right]$, ${\bf X}={\bf e}_k$,
${\bf Q}=-\left[1 \quad {\bf 0}\right]$, and $ {\bf
  \Gamma}_k=\begin{bmatrix} {\bf D}_k\\ {\bf 0} \end{bmatrix}$.

\vspace{2mm}
\hspace{-4.5mm}
Application of {\em Lemma 1} to \eqref{nform22} and \eqref{conref1},
as in Sec. \ref{sec41}, leads to the following SDP formulation of the
first sub-problem:

\vspace{-2mm}
{\small
\begin{eqnarray}
\label{sd22a}
\min_{{\bf B},{\bf G},r,\beta}  && \hspace{-4mm} r\\
\mbox{subject to}&& \hspace{-4mm}  \left[\begin{array}{ccc}
\eta_k-\beta_k & \begin{bmatrix}
({\bf D}_k{\widehat {\bf h}}_k-{\bar{\bf g}}_k\\
  \sigma_n{\bf c}_k\end{bmatrix}^H & {\bf  0}\\
 \begin{bmatrix}{\bf D}_k{\widehat {\bf h}}_k-{\bar{\bf g}}_k \\
 \sigma_n{\bf c}_k\end{bmatrix}  &
        {\bf I}
 & -\delta_k{\bf \Gamma}_k\\
{\bf 0} & -\delta_k{\bf \Gamma}_k^H & \beta_k{\bf I}\end{array}\hspace{-2mm}\right]\succeq{\bf 0}, \nonumber \\ 
& & \hspace{4cm} \beta_k \geq 0, \,\,\,\, \forall k,\nonumber\\
&&\|{\bf b}\| \leq r. \nonumber
\end{eqnarray}
}

\vspace{-2mm}
\hspace{-4mm}
Following a similar approach, it is easy to see that the second
sub-problem can be formulated as the following convex optimization
program:

\vspace{-2mm}
{\small
\begin{eqnarray}
\label{sd22rx}
\min_{{\bf c},s,\mu}  && s_k\\
\mbox{subject to}&&  \left[\begin{array}{ccc}
s_k-\mu_k & \begin{bmatrix}
({\bf D}_k{\widehat {\bf h}}_k-{\bar{\bf g}}_k\\
  \sigma_n{\bf c}_k\end{bmatrix}^H & {\bf  0}\\
 \begin{bmatrix}{\bf D}_k{\widehat {\bf h}}_k-{\bar{\bf g}}_k \\
 \sigma_n{\bf c}_k\end{bmatrix}  &
        {\bf I}
 & -\delta_k{\bf \Gamma}_k\\
{\bf 0} & -\delta_k{\bf \Gamma}_k^H & \mu_k{\bf I}\end{array}\hspace{-2mm}\right]\succeq{\bf 0}, \nonumber \\ 
& & \hspace{4cm} \mu_k \geq 0,  \,\,\,\,\forall k.\nonumber
\end{eqnarray}
}

\vspace{-2mm}
\hspace{-5mm}
The proposed robust MSE-constrained transceiver design algorithm
alternates over both sub-problems. In the next
subsection, we show that this algorithm converges to a limit.

\subsubsection{Convergence}
At the $(n+1)$th iteration, we compute ${{\bf b}}^{n+1}$ and
${{\bf  g}}^{n+1}$ by solving the first sub-problem with fixed
${{\bf c}}^n$. We assume that this sub-problem is feasible,
otherwise the iteration terminates. The solution
of this sub-problem results in ${{\bf b}}^{n+1}$ and
${{\bf g}}^{n+1}$ such that
$f_k({{\bf b}}^{n+1},{{\bf g}}_k^{n+1},{{\bf c}}_k^n)\leq \eta_k, \,\, 1\ \leq k \leq M$, where
\begin{eqnarray}
\label{deff}
f_k & = & \max_{ {{\bf h}}_k: { {\bf h}}_k \in {\cal R}_k}\,\,\epsilon_k.
\end{eqnarray}
Also, the transmit power
$P_{T}^{n+1}=\|{{\bf b}}^{n+1}\|^2 \leq \|{{\bf b}}^n\|^2$.
Solving the second sub-problem in the $n+1$th iteration, we obtain
${{\bf c}}^{n+1}$ such that
\begin{eqnarray}
f_k({{\bf b}}^{n+1},{{\bf g}}_k^{n+1},{{\bf c}}_k^{n+1}) & \leq & f_k({{\bf b}}^{n+1},{{\bf g}}_k^{n+1},{{\bf c}}_k^n).
\end{eqnarray}
Since the transmit power $P_{T}$ is lower-bounded and monotonically
decreasing, we conclude that the sequence $\{P_{T}^n\}$ converges to a
limit as the iteration proceeds.

\subsection{Robust MSE-Balancing Transceiver Design}
\label{sec43}
We next consider the problem of MSE-balancing under a constraint on the
total BS transmit power in the presence of CSI imperfections. When the
CSI is known perfectly, the problem of MSE-balancing can be written as
\begin{eqnarray}
\label{balpr}
\min_{{\bf B},{\bf G},{\bf C}} &&\max_{k} \, \epsilon_k \\
\text{subject to} && \text{tr}({\bf B}{\bf B}^H) \leq P_{max}. \nonumber
\end{eqnarray}
This problem is related to the SINR-balancing problem due to the
inverse relationship that exists between the MSE and SINR. The
MSE-balancing problem in the context of MISO downlink with perfect
CSI has been addressed in \cite{schub04, schub06}. Here, we consider
the MSE-balancing problem in a multiuser MIMO downlink with THP in the
presence of CSI errors. When the CSI is imperfect with NBE model,
this problem can be written as the following convex optimization
problem with infinite
constraints:
\begin{eqnarray}
\label{conref1a}
\min_{{{\bf b}},{{\bf g}},{{\bf c}},r}  && \hspace{-4mm} r\\
\text{subject to} && \hspace{-4mm} \left\|\begin{bmatrix}
{{\bf D}}_k{{\bf h}}_k-{{\bar{\bf
      g}}}_k\nonumber\\
\sigma_n{ {\bf c}}_k
\end{bmatrix}\right\|^2 \leq r, \,\, \forall \, {{\bf h}}_k \in {\cal R}_k,\,\, 1 \leq k \leq M ,\nonumber\\
&&\|{{\bf b}}\|<\sqrt{P_{max}}\nonumber.
\end{eqnarray}
An iterative algorithm, as
in Sec. \ref{sec42},  which involves the solution of two sub-problems  in each iteration can be adopted to solve the above problem. Transforming
the first constraint into an LMI by Schur Complement Lemma,
and then applying
Lemma 1, we can see that the first sub-problem which involves
optimization over ${{\bf b}}$ and ${{\bf g}}$,
for fixed ${{\bf c}}$, is equivalent to the following convex
optimization problem:
\begin{eqnarray}
\label{rbprd1}
\min_{{\bf b },{\bf g}, r,\mu} && \hspace{-5.5mm} r\\
\text{subject to} && \hspace{-5.5mm}
\left[\begin{array}{ccc}
r-\mu_k & \begin{bmatrix}
({\bf D}_k{\widehat {\bf h}}_k-{\bar{\bf g}}_k\\
  \sigma_n{\bf c}_k\end{bmatrix}^H & {\bf  0}\\
 \begin{bmatrix}{\bf D}_k{\widehat {\bf h}}_k-{\bar{\bf g}}_k \\
 \sigma_n{\bf c}_k\end{bmatrix}  &
        {\bf I}
 & -\delta_k{\bf \Gamma}_k\\
{\bf 0} & -\delta_k{\bf \Gamma}_k^H & \mu_k{\bf I}\end{array}\hspace{-2mm}\right]\hspace{-1mm}\succeq{\bf 0}, \nonumber \\ 
& & \hspace{4cm} 1 \leq k \leq M, \nonumber\\
&&\|{{\bf b}}\|<\sqrt{P_{max}}.\nonumber
\end{eqnarray}}
The second sub-problem which involves
optimization over ${{\bf c}}$,
for fixed ${{\bf b}}$ and ${\bf g}$ can be reformulated as in \eqref{s2def}.
By similar arguments as in the MSE-constrained problem, we can see
that this iterative algorithm converges to a limit.

\section{Simulation Results}
\label{sec5}
In this section, we present the performance of the proposed robust
THP transceiver algorithms, evaluated through simulations. We compare
the performance of the proposed robust designs with those of the
non-robust transceiver designs as well as robust linear transceiver
designs reported in the recent literature. The channel is assumed
to undergo flat Rayleigh fading, i.e., the elements of the channel matrices
${\bf H}_k,\,\,1 \leq k \leq M$, are assumed to be independent and
identically distributed (i.i.d) complex Gaussian with zero mean and unit
variance. The noise variables at each antenna of each user terminal are
assumed to be zero-mean complex Gaussian. In all the simulations, all
relevant matrices are initialized as unity matrices. The convergence
threshold is set as $10^{-3}$.

\begin{figure}
\includegraphics[width=3.45in]{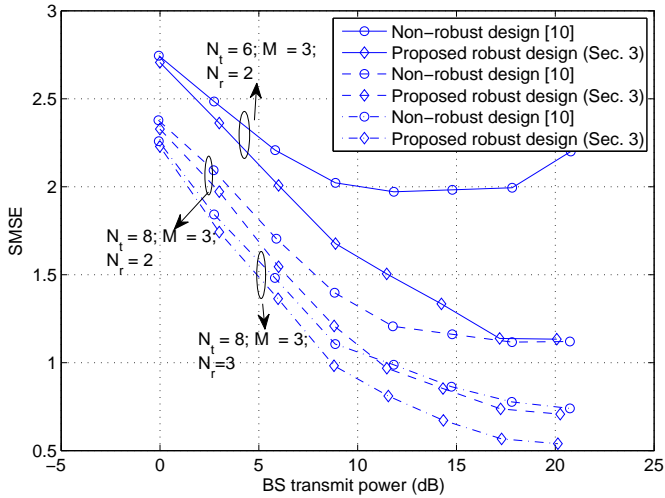}
\caption{
SMSE versus BS transmit power ($P_{T}=\|{\bf B}\|^2_F$) performance
of the proposed robust design in Sec. \ref{sec3} for the SE model.
$N_t=8,\,6$, $M=3$, $N_{r_1}=N_{r_2}=N_{r_3}=2$, $L_1=L_2=L_3=2$,
$\sigma_n^2=1$, $\sigma_E^2=0.1$. {\em Proposed robust design in Sec.
\ref{sec3} outperforms the non-robust design in \cite{mezgh06_2}.}
}
\label{smse_pt}
\end{figure}

\begin{figure}
\includegraphics[width=3.45in]{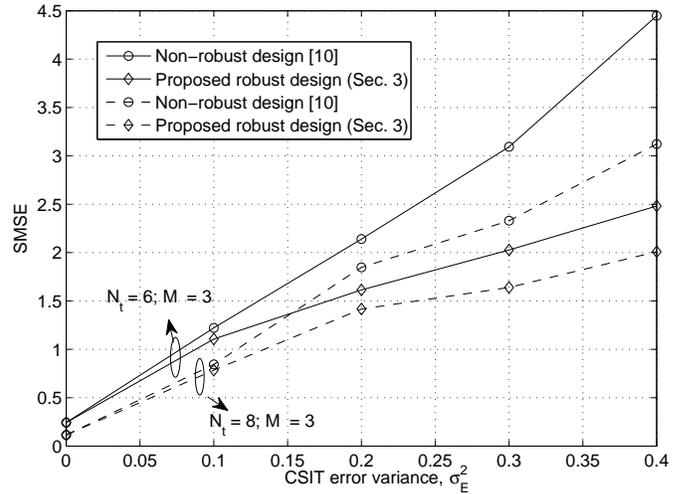}
\caption{
SMSE versus CSIT error variance ($\sigma_E^2$) performance of
the proposed robust design in Sec. \ref{sec3} for the SE model. $N_t=6,\,8$,
$M=3$, $N_{r_1}=N_{r_2}=N_{r_3}=2$, $L_1=L_2=L_3=2$, $P_{max}=15 \,\,\text{dB}$,
$\sigma_n^2=1$. {\em Larger the value of $\sigma_E^2$, higher is the
performance improvement due to the proposed design in Sec. \ref{sec3}
compared to the non-robust design in \cite{mezgh06_2}.}
}
\label{smse_sig}
\end{figure}

First, we consider the performance of the robust transceiver design
presented in Sec. \ref{sec3} for the stochastic CSIT error model. We
consider a system with the BS transmitting $L=2$ data streams each
to $M=3$ users. In Fig. \ref{smse_pt}, we present the simulated
SMSE performances of the proposed robust design and those of the
non-robust design proposed in \cite{mezgh06_2} for different numbers
of transmit antennas at the BS and receive antennas at the user terminals.
Specifically, we consider three configurations: $i)$ $N_t=6$, $N_r=2$,
$ii)$ $N_t=8$, $N_r=2$, and $iii)$ $N_t=8$, $N_r=3$. We use
$\sigma_{E}^2=0.1$ in all the three configurations. From Fig.
\ref{smse_pt}, it can be observed that, in all the three configurations,
the proposed robust design clearly outperforms the non-robust design in
\cite{mezgh06_2}. Comparing the results for $N_t=6$ and $N_t=8$, we find
that the difference between the non-robust design and the proposed robust
design decreases when more transmit antennas are provided. A similar
effect is observed for increase in number of receive antennas for fixed
number of transmit antennas. It is also found that the difference between
the performance of these algorithms increase as the SNR increases. This
is observable in (\ref{e11}), where the second term shows the effect of
the CSIT error variance amplified by the transmit power. In Fig.
\ref{smse_sig}, we illustrate the SMSE performance as a function of
different channel estimation error variances, $\sigma^2_E$, for similar
system parameter settings as in Fig. \ref{smse_pt}. In Fig. \ref{smse_sig}
also, we observe that the proposed robust design performs better than the
non-robust design in the presence of CSIT error; larger the estimation
error variance, higher is the performance improvement due to
robustification in the proposed algorithm.

\begin{figure}
\includegraphics[width=3.45in]{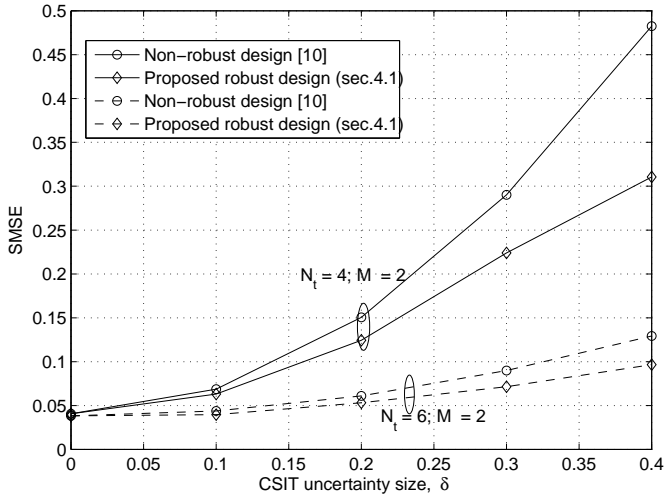}
\caption{
SMSE versus CSIT uncertainty size ($\delta$) performance of
the proposed robust design in Sec. \ref{sec41} for the NBE model.
$N_t=6,\,4$, $M=2$, $N_{r_1}=N_{r_2}=N_{r_3}=2$, $L_1=L_2=L_3=2$,
$\delta_1=\delta_2=\delta$, $P_{max}=15 \,\,\text{dB}$, $\sigma_n^2=0.1$.
{\em Proposed robust design in Sec. \ref{sec41} performs better than
the design in \cite{mezgh06_2}.} 
}
\label{smse_dl}
\end{figure}

\begin{figure}
\includegraphics[width=3.45in]{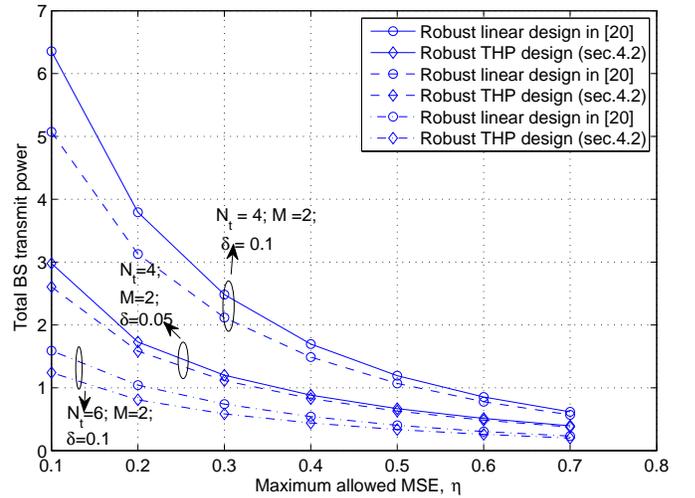}
\caption{
Total BS transmit power ($P_{T}=\|{\bf B}\|^2_F$) required as a
function of maximum allowed MSE at the user terminals ($\eta_1=\eta_2=\eta$)
in the proposed robust design in Sec. \ref{sec42} for the NBE model.
$N_t=4,\,6$, $M=2$, $N_{r_1}=N_{r_2}=2$, $L_1=L_2=2$, CSIT uncertainty
range $\delta_1=\delta_2=\delta= 0.05, 0.1$. {\em Proposed robust THP
transceiver design in Sec. \ref{sec42} requires lesser BS transmit power
to meet the MSE constraints at the user terminals than the robust linear
transceiver design in \cite{vucic08}. }
}
\label{pt_mse}
\end{figure}

Next, we present the performance of the robust transceiver designs
proposed in Sec. \ref{sec4} for the norm-bound model of CSIT error.
Figure \ref{smse_dl} shows the SMSE performance of the proposed design
in Sec. \ref{sec41} as a function of the CSIT uncertainty size, $\delta$,
for the following system settings: $N_t=6,\,4$, $M=2$,
$N_{r_1}=N_{r_2}=N_{r_3}=2$, $L_1=L_2=L_3=2$, $\delta_1=\delta_2=\delta$,
$P_{max}=15 \,\,\text{dB}$, and $\sigma_n^2=0.1$. It is seen that the proposed
design in Sec. \ref{sec41} is able to provide improved performance compared
to the non-robust transceiver design in \cite{mezgh06_2}, and this
improvement gets increasingly better for increasing values of the CSIT
uncertainty size, $\delta$. In Fig. \ref{pt_mse}, we illustrate the
performance of the robust MSE-constrained design proposed in Sec.
\ref{sec42} for the following set of system parameters: $N_t=4,\,6$,
$M=2$, $N_{r_1}=N_{r_2}=2$, $L_1=L_2=2$, and
$\delta_1=\delta_2=\delta=0.05,0.1$. We plot
the total BS transmit power, $P_{T}=\|{\bf B}\|_F^2$, required to achieve
a certain maximum allowed MSE at the user terminals, $\eta_1=\eta_2=\eta$.
As expected, as the maximum allowed MSE is increased, the required total
BS transmit power decreases. For comparison purposes, we have also shown
the plots for the robust {\em linear} transceiver design presented in
\cite{vucic08} for the same NBE model. It can be seen that the proposed
THP transceiver design needs lesser total BS transmit power than the
robust linear transceiver in \cite{vucic08} for a given maximum allowed
MSE. The improvement in performance over robust linear transceiver is
more when the maximum allowed MSE is small.

\begin{figure}
\includegraphics[width=3.45in]{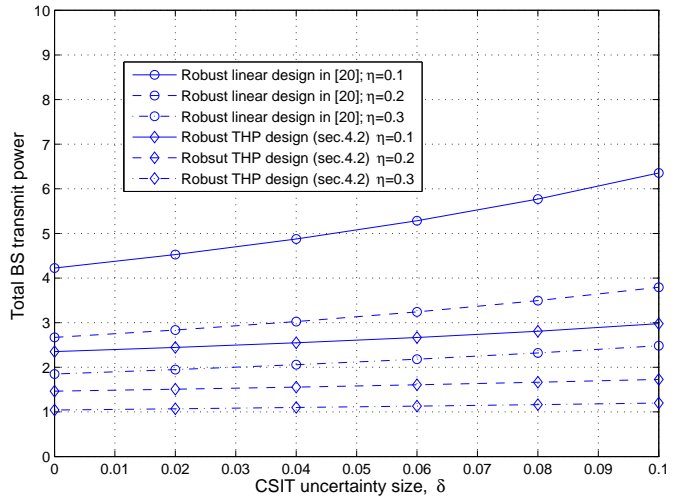}
\caption{
Total BS transmit power ($P_{T}=\|{\bf B}\|^2_F$) required as a
function of CSIT uncertainty size, $\delta$, to meet MSE constraints in
the proposed robust design
in Sec. \ref{sec42} for the NBE model. $N_t=4$, $M=2$, $N_{r_1}=N_{r_2}=2$,
$L_1=L_2=2$, maximum allowed MSEs $\eta_1=\eta_2=\eta= 0.1, 0.2, 0.3$, and
$\delta_1=\delta_2=\delta$. {\em Proposed robust THP transceiver design
requires lesser BS transmit power than the robust linear transceiver design
in \cite{vucic08}.}
}
\label{pt_dl}
\end{figure}

\begin{figure}
\includegraphics[width=3.45in]{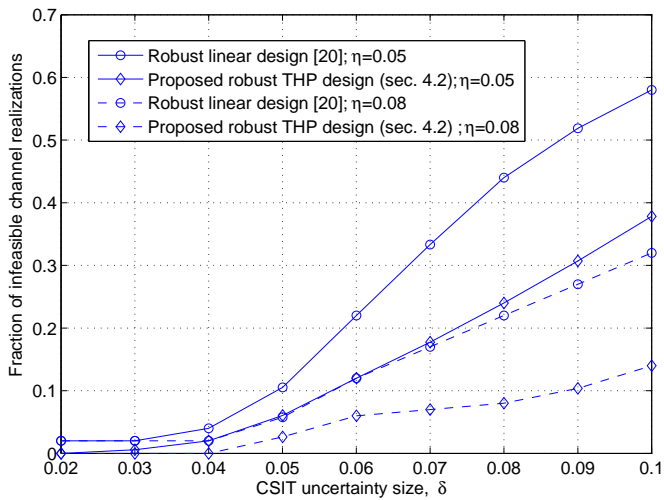}
\caption{
Fraction of infeasible channel realizations for different values
of CSIT uncertainty size $\delta_1=\delta_2=\delta$ in the proposed robust
design in Sec. \ref{sec42} for the NBE model. $N_t=4$, $M=2$,
$N_{r_1}=N_{r_2}=2$, $L_1=L_2=2$, and $\eta_1=\eta_2=\eta=0.05,0.08$.
{\em Proposed robust THP transceiver design in Sec. \ref{sec42} has
lesser infeasible channel realizations than the robust linear design in
\cite{vucic08}.}
}
\label{infs}
\end{figure}

\begin{figure}
\includegraphics[width=3.45in]{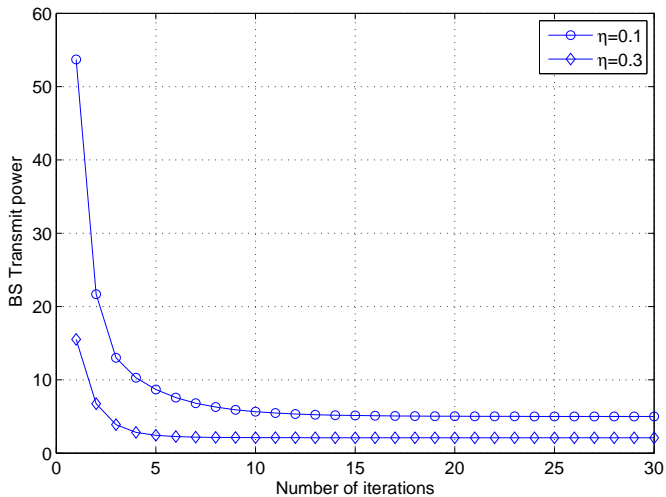}
\caption{
Convergence behavior of the proposed robust THP transceiver design 
in Sec. \ref{sec42}. CSIT uncertainty size $\delta_1=\delta_2=\delta=0.1$.
$N_t=4$, $M=2$, $N_{r_1}=N_{r_2}=2$, $L_1=L_2=2$, and 
$\eta_1=\eta_2=\eta=0.1,0.3$.
}
\label{convr}
\end{figure}

Further, in Fig. \ref{pt_dl}, we present the total BS transmit
power required to meet MSE constraints at user terminals for different
values of CSIT uncertainty size $\delta_1=\delta_2=\delta$, for $N_t=4$,
$M=2$, $N_{r_1}=N_{r_2}=2$, $L_1=L_2=2$, and maximum allowed MSEs
$\eta_1=\eta_2=\eta= 0.1, 0.2, 0.3$. As can be seen from Fig. \ref{pt_dl},
the proposed robust THP transceiver design in Sec. \ref{sec42} meets the
desired MSE constraints with much less BS transmit power compared to the
robust linear transceiver in  \cite{vucic08}. We note that infeasibility
of robust transceiver design for certain realizations of channels is
an issue in robust designs with MSE constraints. In Fig. \ref{infs},
we show the performance of the proposed THP transceiver design in Sec.
\ref{sec42}, in terms of the fraction of channel realizations for which
the design is infeasible for $N_t=4$, $M=2$, $N_{r_1}=N_{r_2}=2$, and
different values of $\delta$ and $\eta$. It is seen that that the
fraction of infeasible channel realizations in case of the proposed
THP transceiver is much less compared to that in case of the linear
transceiver in \cite{vucic08}. For example, for CSIT uncertainty size
$\delta_1=\delta_2=\delta=0.08$ and user MSE $\eta_1=\eta_2=\eta=0.05$,
the linear transceiver design fails to produce a feasible solution in
about $44\%$ cases, whereas the proposed THP transceiver design fails
only in about $24\%$ cases. In Fig. \ref{convr}, we show the
convergence behavior of the proposed design. The number of iterations
to converge depends on the MSE constraints. Stricter MSE constraints
lead to larger number of iterations to converge.
For example, when the MSE constraint is $\eta=0.3$, the algorithm
converges in around 6 iterations. Whereas, for $\eta=0.1$, it takes
around 12 iterations to converge.

\begin{figure}
\includegraphics[width=3.45in]{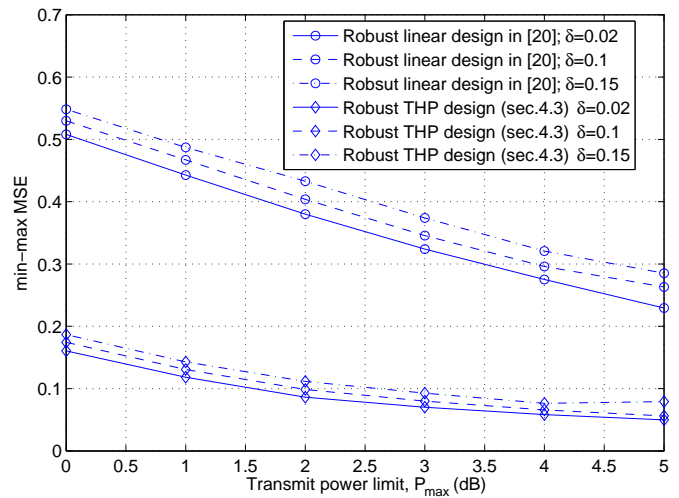}
\caption{
Min-max MSE versus total BS transmit power limit, $P_{max}$,
in the proposed
robust design in Sec. \ref{sec43} for the NBE model. $N_t=4$, $M=2$,
$N_{r_1}=N_{r_2}=2$, $L_1=L_2=2$, $\delta_1=\delta_2=\delta=0.02,0.1,0.15$.
{\em Proposed robust THP transceiver design in Sec. \ref{sec43} performs
better than the robust linear design in \cite{vucic08}.}
}
\label{mse_pt}
\end{figure}

Finally, in Fig. \ref{mse_pt}, we present the performance of the proposed
robust MSE-balancing transceiver design in Sec. \ref{sec43} for  $N_t=4$,
$M=2$, $N_{r_1}=N_{r_2}=2$, $L_1=L_2=2$,
$\delta_1=\delta_2=\delta=0.02,0.1,0.15$. The min-max MSE plots as a
function of total BS transmit power constraint are shown. The
corresponding plots for the robust linear transceiver in \cite{vucic08}
are also shown. The results in Fig. \ref{mse_pt} show that, for the same
BS transmit power constraint, the proposed robust design in Sec. \ref{sec43}
achieves lower min-max MSE compared to the robust linear transceiver
design in \cite{vucic08}.

\section{Conclusions}
\label{sec6}
We proposed robust THP transceiver designs that jointly optimize
the THP precoder and receiver filters in multiuser {\em MIMO} downlink
in the presence of {\em imperfect CSI at the transmitter}. We considered
these transceiver designs under SE and NBE models for CSIT errors.
For the SE model, we proposed a minimum SMSE transceiver design.
For the NBE model, we proposed three robust designs, namely, minimum
SMSE design, MSE-constrained design, and MSE-balancing design. We
presented iterative algorithms to solve these robust design problems.
The iterative algorithms involved solution of sub-problems, which
have either analytical solutions or can be formulated as convex
optimization problems that can be solved efficiently. Through
simulation results we illustrated the superior performance of the
proposed robust designs compared to non-robust designs as well as
robust linear transceiver designs that have been reported recently
in the literature.

\begin{IEEEbiography}[]{P. Ubaidulla}
received the B.Tech degree in Electronics and
Communication Engineering from the National Institute of
Technology (NIT), Calicut in 1997, and the M.E degree in
Communication Engineering from NIT, Trichy in 2001.
From 1998 to 1999, he was with Hindustan Petroleum
Corporation as an engineer. From 2001 to 2005, he was
with Central Research Laboratory, Bharat Electronics
Limited, Bangalore, India, working on signal processing
algorithms for radar and  sonar applications. Since 2005
he has been working towards the Ph.D degree in wireless
communications at the Department of Electrical Communication
Engineering, Indian Institute of Science, Bangalore, India.
His current research interests are in multiuser MIMO
communications and robust optimization.
\end{IEEEbiography}

\begin{IEEEbiography}[]{A. Chockalingam}
was born in Rajapalayam, Tamil Nadu, India. He
received the B.E. (Honors) degree in Electronics and Communication
Engineering from the P. S. G. College of Technology, Coimbatore,
India, in 1984, the M.Tech degree with specialization in satellite
communications from the Indian Institute of Technology, Kharagpur,
India, in 1985, and the Ph.D. degree in Electrical Communication
Engineering (ECE) from the Indian Institute of Science (IISc),
Bangalore, India, in 1993. During 1986 to 1993, he worked with the
Transmission R \& D division of the Indian Telephone Industries
Limited, Bangalore. From December 1993 to May 1996, he was a
Postdoctoral Fellow and an Assistant Project Scientist at the
Department of Electrical and Computer Engineering, University of
California, San Diego. From May 1996 to December 1998, he served
Qualcomm, Inc., San Diego, CA, as a Staff Engineer/Manager in the
systems engineering group. In December 1998, he joined the faculty
of the Department of ECE, IISc, Bangalore, India, where he is a
Professor, working in the area of wireless communications and 
networking.

Dr. Chockalingam is a recipient of the Swarnajayanti Fellowship from
the Department of Science and Technology, Government of India. He
served as an Associate Editor of the IEEE Transactions on Vehicular
Technology from May 2003 to April 2007. He currently serves as an
Editor of the IEEE Transactions on Wireless Communications. 
He served as a Guest Editor for the IEEE JSAC Special Issue on
Multiuser Detection for Advanced Communication Systems and Networks.
He is a Fellow of the Institution of Electronics and Telecommunication 
Engineers, and a Fellow of the Indian National Academy of Engineering.
\end{IEEEbiography}

\vfill
\end{document}